# Tunable bipolar optical interactions between guided lightwaves

Mo Li[1], W. H. P. Pernice[1], H. X. Tang[1]

[1]*Departments of Electrical Engineering, Yale University, New Haven, CT 06511, USA*

**The optical binding forces between guided lightwaves in dielectric waveguides can be either repulsive or attractive. So far only attractive force has been observed. Here we experimentally demonstrate a bipolar optical force between coupled nanomechanical waveguides. Both attractive and repulsive optical forces are obtained. The sign of the force can be switched reversibly by tuning the relative phase of the interacting lightwaves. This tunable, bipolar interaction forms the foundation for the operation of a new class of light force devices and circuits.**

State-of-the-art advances in nanophotonics allow light to be highly concentrated in nanoscale waveguides or resonators with high refractive index contrast. In closely spaced devices the coupling between the guided lightwaves gives rise to an optical force known as the "optical binding force". According to recent theoretical predictions[1-4], the polarity of this force is either attractive or repulsive depending on the relative phase between the interacting lightwaves.

Recently, attractive-type optical force was experimentally demonstrated in suspended photonic structures[5], whereas repulsive optical force has not been confirmed yet. However, repulsive optical force is indispensible for achieving reversible force control in a wide range of applications such as tunable photonic devices[6], optomechanical signal processing and all-optical switching[7]. In addition, future developments of cavity optomechanics[8] will benefit from an optomechanical force of tunable sign.

The optical force is a form of electrodynamic interaction that is of classical origin and intrinsic to Maxwell equations. This classical force is closely related to the electrodynamic force arising from vacuum quantum fluctuations, known as the Casimir force[9-11]. It was recently demonstrated that the Casimir force can be bipolar if the dielectric properties of the interacting media are carefully chosen[12].

The challenge of obtaining the repulsive optical force lies in achieving independent phase control of interacting lightwaves in coupled systems. In this work, we demonstrate repulsive optical force between lightwaves guided in two coupled optical waveguides. We further demonstrate in-situ bipolar tuning of this force, from repulsive to attractive and vice versa, by adjusting the relative phase of the optical modes. This unified picture of optical forces represents a significant advance from the previously observed unipolar optical force[4] and verifies recent theories[1-4,13].

Considering two proximate dielectric waveguides separated by gap $g$ as depicted in Fig. 1a, propagating lightwaves in each waveguide interact through the excitation of dipole oscillations. The optically excited dipoles in one waveguide interact with the evanescent field of the other waveguide and generate a force which is dependent on the separation of the waveguides and the relative phase of optical modes. When the modes are in phase ($\phi = 0$), the optical force is always attractive; when the modes are out of phase ($\phi = \pi$), the resulting force is repulsive when the separation is larger than a critical value $g_c$ as shown in Fig. 1b. Here the calculation is carried out for two identical single-mode silicon waveguides with a cross-section of 300×220 nm$^2$. Beyond $g_c$, when the phase difference is varied, the sign of the force oscillates between positive and negative values as shown in Fig. 1c. In Fig. 1d, we plot the calculated amplitude of the force versus the separation $g$ between the waveguides and the relative phase $\phi$ between the optical modes.

When the two waveguides are free-standing, this interaction force causes mechanical displacement of the beams: the attractive optical force pulls the two waveguides closer while the repulsive force pushes them apart, as schematically shown in the insets of Fig.1b. By measuring the amplitude of the relative nanomechanical motion in response to optical excitation, the magnitude of the force between the waveguides can be precisely quantified. The sign of the force is determined from the phase of the vector response.

We employ a Mach-Zehnder (MZ) cascade configuration to control the phase difference between the incoming lightwaves (Fig. 2a) by adjusting the input wavelength. The MZ is separated into two symmetric halves, joined by a central coupling region (Fig. 2b). In each half of the MZ, the path length of the top arm is 90 μm longer than the bottom arm. In the 12 μm long coupling region, the top and bottom waveguides are brought together to allow the lightwaves to overlap. As shown in the scanning electron microscope (SEM) image in Fig. 2b, the coupled waveguides are suspended from the substrate so that they are free to move in-plane when a interaction force arises between them. The unsuspended parts of the waveguides have a cross-section of 500nm × 220nm, whereas in the suspended regime they are tapered down to 300nm × 220nm to maximize in-plane optomechanical interactions. In order to provide precise control of the beam length and to attain rigid mechanical support, the waveguides are joined through a pair of 2×2 photonic crystal waveguide couplers. The couplers are optimized for low-loss transmission and cause minimal perturbation of the waveguide modes. (Details of waveguide and coupler designs are presented in the supplementary materials.)

After passing through the first half of the MZ, the lightwaves in the top and bottom arms gain a phase difference of $\Delta\phi = 2\pi \cdot n_{eff} \Delta L / \lambda$. Here $n_{eff}$ is the effective index of the waveguide, $\Delta L$ is the path length difference and $\lambda$ is the wavelength. After leaving the coupling region the optical modes gain the same $\Delta\phi$ in the second half of the MZ before they are combined at the output. Clear interference patterns can be measured in the transmission spectrum of the device, as shown in Fig. 2c ($g$=400nm) and Fig. 2e ($g$=100nm). The separation $g$ between the two waveguides in the coupling region impacts the phase properties of the MZ. When $g$ is large, the coupling between the two waveguides is weak so that the phase difference of the lightwaves in both waveguides is maintained in the coupling region, except for a small amount of phase shift due to the coupling. However, when $g$ is small, the optical modes in the two waveguides will couple strongly. As a result the two halves of the MZ are decoupled and the transmission spectrum is similar to that of two short MZs joined in series, with 90 µm path length difference in each. These two situations are schematically shown in Fig. 2e. In Fig. 2c, the periodicity of the fringes is ~3nm at 1550 nm wavelength, corresponding to a total path length difference of 180 µm and a group index of ~4 in the silicon waveguides. From the observed good extinction ratio it is clear that the optical field intensity in the two arms is well balanced. The situation is changed for the strongly coupled case shown in Fig. 2d, where every other of the interference fringes in Fig. 2c disappears. Comparing the two spectra allows us to determine the order (even or odd) of each interference fringe, as labeled in Fig. 2c and d. A detailed analysis of the coupling effect using coupled mode theory (CMT) is described in supplementary materials.

In the following we focus on the weak coupling regime where the optical phase difference in the two waveguides can be well controlled. At the peaks of the transmission spectrum (Fig. 2c), the total phase difference at the output of the MZ cascade satisfies the condition $2\Delta\phi = 2M\pi$, where $M$ is an integer. Therefore due the symmetry, in each half of the MZ, the phase difference is $\Delta\phi = M\pi$. The orders of adjacent peaks in the interference pattern are $M=2N$ and $M=2N+1$, respectively. At the central coupling region, the optical modes are in-phase (symmetric) if $\Delta\phi = 2N\pi$ and out-of-phase (anti-symmetric) if $\Delta\phi = (2N+1)\pi$. By tuning the input wavelength we can thus selectively excite the in-phase or out-of-phase optical modes of the central coupling region. This allows us to precisely control the sign of the interaction force, which is negative for the symmetric mode and positive for the anti-symmetric mode.

The optically excited mechanical responses of the suspended waveguides are then used to measure the lightwave interaction forces. An actuating light is amplitude modulated, generating dynamic interaction forces between the waveguides. A probing light of fixed wavelength, offset from the actuating light, is applied at the maximal slope (the quadrature point) of the even order interference fringe to detect the nanomechanical motion of the waveguides using an interferometric method[5]. As shown in Fig. 3a, two prominent resonance peaks are observed at frequencies of 17.05 and 18.64 MHz, with quality factors of 5300 and 5400, respectively. These correspond to the in-plane fundamental vibration modes of each suspended waveguide; the very weak out-of-plane fundamental resonances are found at lower frequencies of ~11 MHz. Because the separation between the waveguides and substrate are intentionally large (~1 µm), the out-of-plane gradient optical force is insignificant in this case[5]. The difference of the observed resonance frequencies of the two waveguides is attributed to the unbalanced mechanical clamping and fabrication imperfection (see supplementary materials).

To quantify the interaction force between the waveguides, the resonance of either waveguide can be used, since we only measure their relative motion. Here, we focus on the waveguide with lower resonance frequency. At this frequency, the higher frequency waveguide is off-resonance and can be treated as quasi-stationary. In the linear regime, the waveguide's resonating amplitude is proportional to the modulation amplitude of the actuating light power (inset, Fig. 3a). The resonator's phase response to an external harmonic driving force can be used to determine the sign of the driving force unambiguously. This can be seen from the simple response function of a forced resonator[14]:

$$\delta g(\omega) = g(\omega) - g_0 = \frac{F}{m(\omega_0^2 - \omega^2 + i\omega_0\omega/Q)} \qquad (1),$$

where $\omega$ is the angular frequency, $\omega_0$ is the resonance frequency, $F$ is the driving force, $m$ is the mass, and $Q$ is the quality factor. A positive sign of $F$ indicates that the force pushes the beams apart ($\delta g > 0$); a negative sign of $F$ indicates that the force pulls the beam closer ($\delta g < 0$). By matching above expression to the observed resonance curves in both amplitude and phase, we can determine the sign of the force $F$. The observed optical signal is related to the displacement $\delta g$ by a transduction factor $G = (\Delta T / T_0)/\delta g$ which has a constant positive value of 2.2 µm$^{-1}$ when the probing light is fixed at the quadrature of an even order interference fringe (1538.9 nm in Fig. 2c; the calibration procedure is described in supplementary materials.). However, the exact value and sign of $G$ will not affect the determination of the sign of the measured force $F$ because the detected signal is normalized to this factor in all the measurements. According to equation (1), when $F$ is positive, the resonator's phase changes from 0 to $-\pi$ when the actuation frequency is swept through the resonance frequency from low to high frequencies. When $F$ turns negative, the phase will be shifted by $\pi$ and thus evolve from $\pi$ to 0. Such an effect is observed and shown in Fig. 3b when the actuation laser wavelength is tuned from an even order fringe (red) to the adjacent odd order fringe (blue). This $\pi$ phase shift indicates that the direction of the force has changed sign, from attractive (negative) to repulsive (positive). It is more illustrative to plot the resonance response in the real-imaginary complex phase-plane as shown in Fig. 3c. Then the resonance response to the positive (negative) force is represented by a circle in the bottom (top) phase-plane. The plot is obtained by sweeping the actuation wavelength over a wide range (from 1529 to 1562 nm in 330 steps) and recording the normalized complex resonance response of the coupled waveguides. Clearly, the measured resonance responses fall into two different circles in the top and bottom phase plane, corresponding to repulsive (blue) and attractive (red) forces, respectively. In Fig. 3d, we mark the wavelength regions that show repulsive force in blue and the regions that show attractive force in red. It is obvious that when the coupled modes in the two waveguides are symmetric ($\Delta\phi = 2N\pi$), shown as the red fringes, the optical force between the waveguide is attractive; when the coupled mode is anti-symmetric ($\Delta\phi = (2N+1)\pi$), shown as the blue fringes, the optical force changes its sign and becomes repulsive. This is in agreement with the theoretical predictions.

We further quantify the magnitude of the interaction force at different wavelengths using the calibrated displacement sensitivity and the mechanical properties of the beam (see supplementary materials). We measure the resonance amplitude of the waveguide resonator with the actuating light at various wavelengths but constant power and derive the corresponding amplitude of the force on the beam. The result is shown in Fig. 4a. The amplitude of the force shows sinusoidal oscillations; its sign changes between attractive and repulsive as the driving optical wavelength and thus the phase of the optical mode are tuned. We note that the peak amplitudes of the repulsive and attractive force are different, which is expected from the theoretical calculation (shown in Fig. 1b and 1c). At arbitrary phase difference, both symmetric and anti-symmetric eigenmodes of the coupled waveguides are excited. Due to their different propagation constant, the spatial force distribution along the waveguide is not uniform. When the initial phase difference is not 0 or $\pi$, the direction of the force along the waveguides alternates between repulsive and attractive. Thus, the observed net force is the integration of the distributed optical force over the length of the coupled waveguides. This force distribution is modeled using coupled mode theory (CMT). The numerical calculation results for $\Delta\phi=0$, $\pi/2$, and $\pi$ are displayed in Fig. 4b, showing the evolution of the optical mode and the force distribution on the waveguides. In Fig. 4a, the net optical force for arbitrary initial phase difference is calculated and fitted with the data, showing quantitative agreement with the experimental result.

In summary, a bipolar optical force of classical origin has been observed through active control of the phase fronts of coupled optical modes. The sign of the force is tunable in-situ by adjusting the wavelength of the light. This planar optomechanical interaction promises significant advantage for device integration and multiplexing[15]. Full exploitation of this tunable force will enable a new class of light force driven devices and circuits to be realized on a ubiquitous silicon platform in which mechanical components can interact with light routed through waveguides in a layout similar to electronic circuits. Furthermore, the bipolarity of the force is of great interest for cavity optomechanics[16-19], where the sign of an optical spring could be reversed simply by tuning the phase.

**METHODS SUMMARY**

The photonic devices are fabricated on silicon-on-insulator wafers with a 220 nm top silicon layer and a 3 μm thick buried oxide layer. Electron beam lithography and plasma etching are employed to pattern the waveguide structures. Release of the suspended waveguide is achieved by wet chemical etching of the oxide layer using buffered oxide etchant (BOE). Since only the in-plane coupling between waveguides is studied, the etching time is kept long enough to ensure that the release gap is made larger than ~ 1 μm, in order to minimize substrate coupling. We note that in our experiments, while we are able to achieve high precision device repeatability in the photonic domain, the mechanical resonance frequency of the devices show deviations from the expected values in a ±10% range. Several fabrication issues contribute to this distribution including the lithographical repeatability, the residual stress and uncontrolled undercut and clamping.

A pump-probe scheme employing an actuation laser and a probing laser is applied to excite and measure the optical force induced nanomechanical motion of the waveguides. The actuation laser is amplitude modulated with an electro-optical modulator. The response is measured in the transmitted signal, after passing through a band-pass inline tunable filter (1nm - 3dB bandwidth) centered at the probing wavelength, using a high-speed photodetector (New Focus 1811) and a vector network/spectrum analyzer.

**Figures**

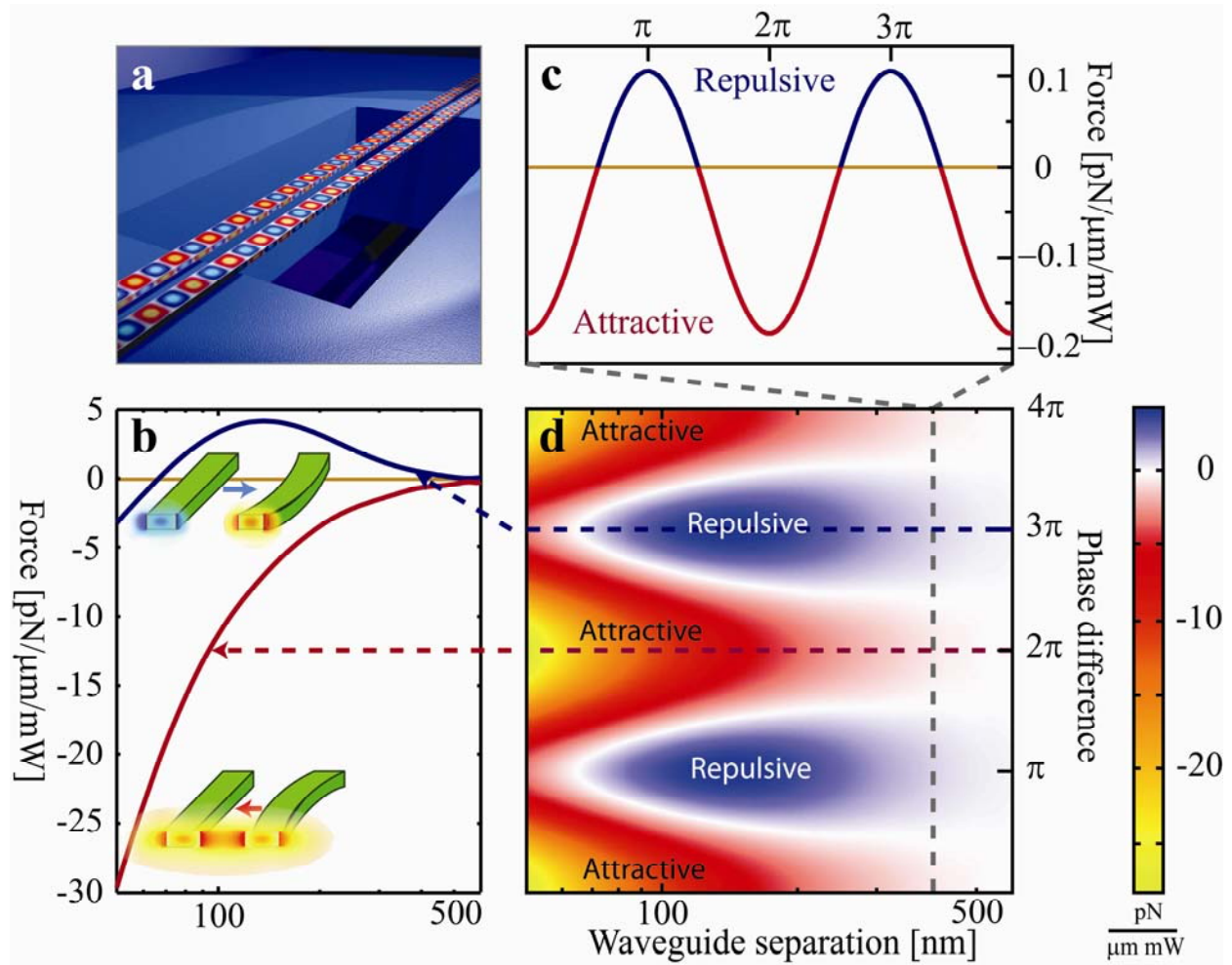

**Figure 1. Optical interaction between two coupled waveguides**. **a**, Schematic illustration of two suspended waveguides. **b,** The optical force between the waveguides as a function of their separation for symmetric and anti-symmetric coupled modes. **c,** The amplitude of the optical force oscillates with the relative phase between the modes in each waveguide at a fixed separation of 400 nm. **d,** Numerical simulation result of the optical force as a function of both the waveguide separation and the phase difference.

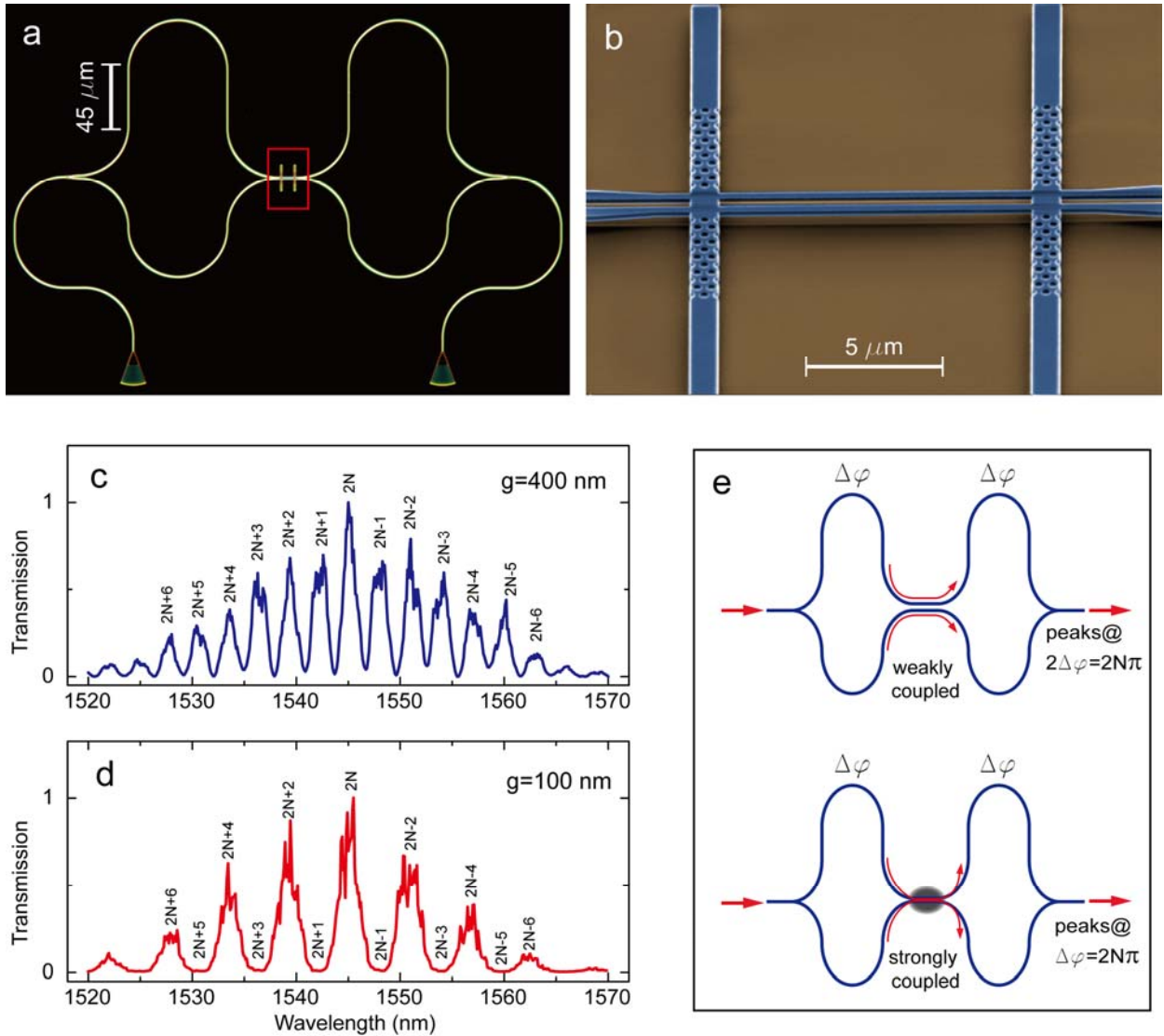

**Figure 2.** Mach-Zehnder (MZ) cascade with the two waveguides coupled at the center. **a**, Optical microscope image (dark field) of the device, showing the input/output grating coupler and the central coupling region. **b**, Scanning electron microscope image of the suspended coupled waveguides with photonic crystal waveguide coupler supporting structures (The region inside the red box in a). **c** and **d,** Measured transmission spectrum of MZ cascade devices in weak ($g$=400nm) and strong ($g$=100nm) coupling situations. The orders of the fringes are labeled in the curve. **e**, Schematics illustrating the situations of weak and strong coupling in the center of the MZ.

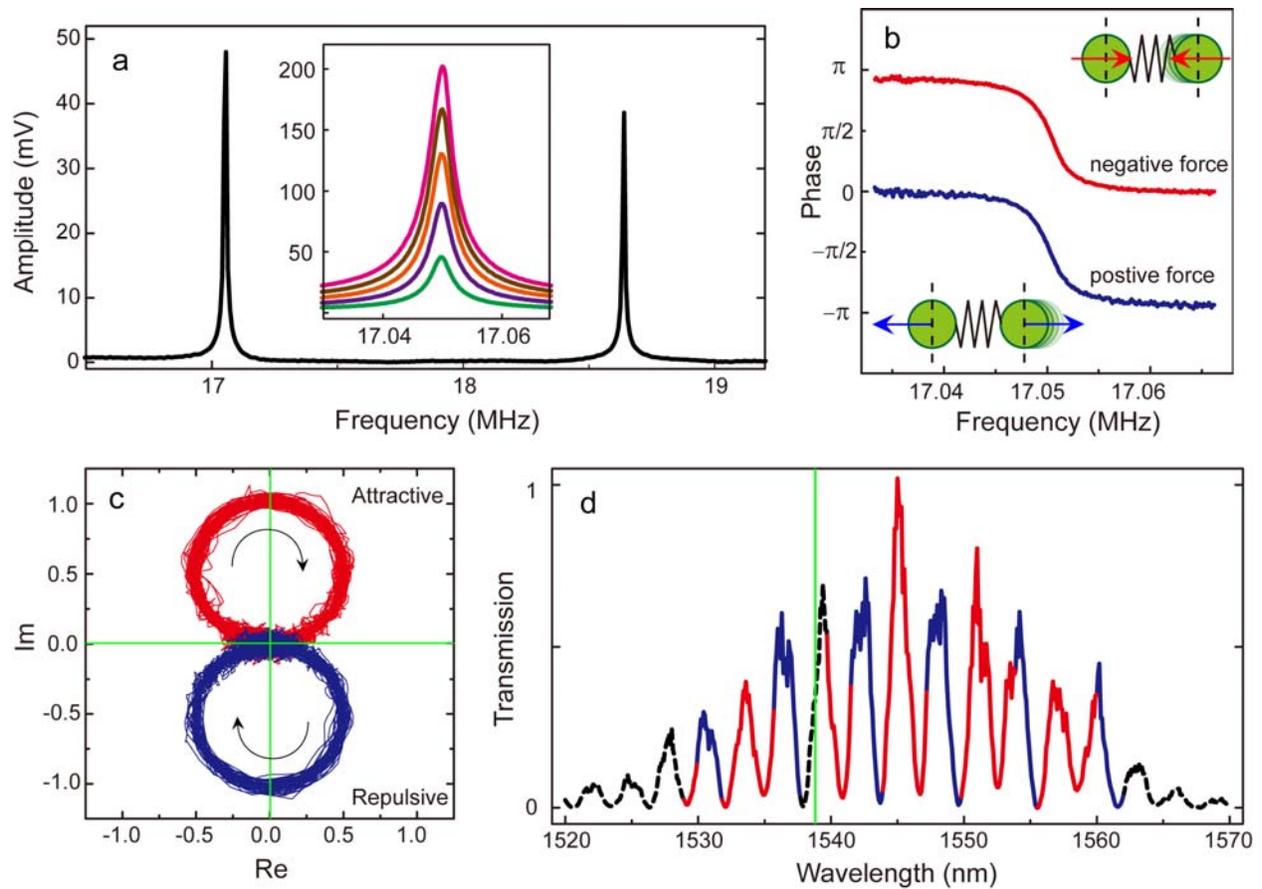

**Figure 3.** In-situ tuning of the interaction force and the waveguides' nanomechanical resonance responses. **a**, Wide range amplitude response showing the two resonance peaks corresponding to each waveguide beam's in-plane resonance mode. Inset: Frequency response of the first peak at various actuation laser modulation level from 100 mV to 500 mV. **b,** The measured phase response of the waveguide's resonance to negative and positive driving forces. Inset schematics illustrate the direction of the forces. **c,** The waveguide's resonance response plotted in phase-space at varying driving laser wavelengths from 1529 to 1562 nm in 330 steps. When the force is repulsive (attractive), the normalized resonance response appears as a circle in the bottom (top) half of the phase plane as shown in blue (red). **d**, The measured transmission spectrum with the wavelength ranges generating repulsive forces colored in blue and attractive forces colored in red. The actuation wavelength is not scanned in the close proximity of the probe wavelength (marked with the green line at 1538.9 nm) to allow for sufficient filtering of the actuation light in the detector.

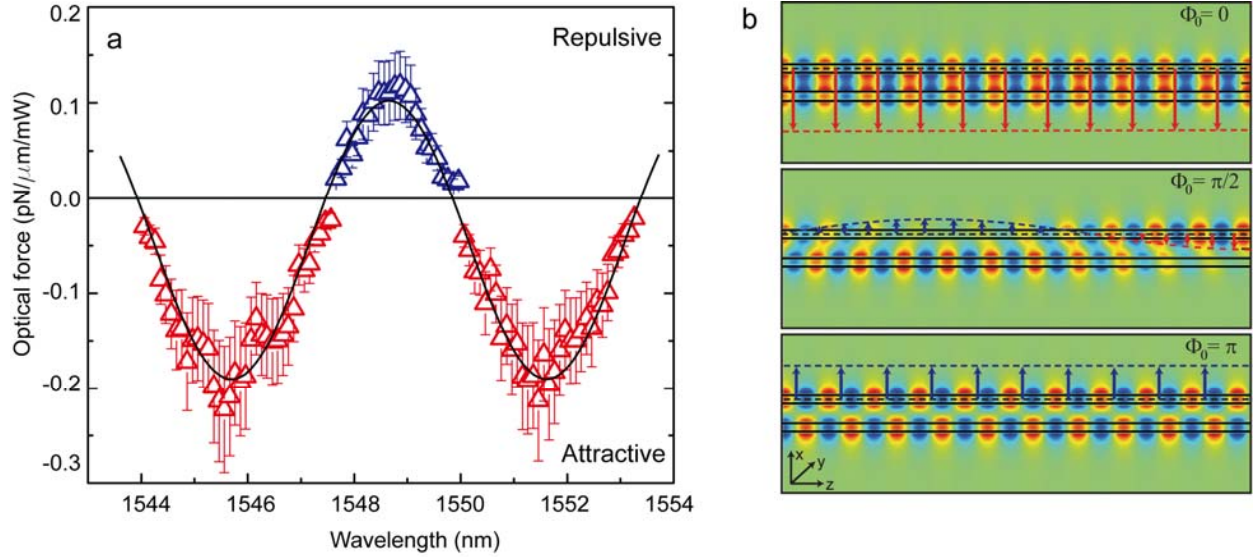

**Figure 4.** Tunable optical force between two coupled waveguides. **a**, The amplitude of the interaction force between the waveguides is measured with varying actuation wavelength. It alternates between repulsive (blue) and attractive (red) sinusoidally with the wavelength, following the prediction from coupled mode theory (black line). Error bars are evaluated from systematic errors in power and displacement calibrations. (See supplementary materials.) **d**, The optical mode in the coupled beam when $\Delta\phi = 0$, $\pi/2$ and $\pi$, corresponding to the maximal attractive force, the medium force and the maximal repulsive force, respectively. The distribution of the force is shown schematically with dotted lines and the direction of the force on the top waveguide is indicated by the arrows.

**Acknowledgement:**

We acknowledge a seedling grant from DARPA/MTO. W.H.P.P. acknowledges support from the Alexander von Humboldt postdoctoral fellowship program. H.X.T. acknowledges support from Yale Institute of Nanoscience and Quantum Engineering and a career award from National Science Foundation. The devices were fabricated at Yale University Microelectronic Center and the NSF-sponsored Cornell Nanoscale Facility.

**Reference:**

1. Povinelli, M. L. *et al.* Evanescent-wave bonding between optical waveguides. *Opt. Lett.* **30**, 3042-3044 (2005).
2. Ng, J., Chan, C. T., Sheng, P. & Lin, Z. F. Strong optical force induced by morphology-dependent resonances. *Opt. Lett.* **30**, 1956-1958 (2005).
3. Mizrahi, A. & Schachter, L. Mirror manipulation by attractive and repulsive forces of guided waves. *Opt. Express* **13**, 9804-9811 (2005).
4. Rakich, P. T., Popovic, M. A., Soljacic, M. & Ippen, E. P. Trapping, corralling and spectral bonding of optical resonances through optically induced potentials. *Nat Photon* **1**, 658-665 (2007).
5. Li, M. *et al.* Harnessing optical forces in integrated photonic circuits. *Nature* **456**, 480-484 (2008).
6. Huang, M. C. Y., Zhou, Y. & Chang-Hasnain, C. J. A nanoelectromechanical tunable laser. *Nature Photon.* **2**, 180-184 (2008).
7. Vlasov, Y., Green, W. M. J. & Xia, F. High-throughput silicon nanophotonic wavelength-insensitive switch for on-chip optical networks. *Nature Photon.* **2**, 242-246 (2008).
8. Kippenberg, T. J. & Vahala, K. J. Cavity opto-mechanics. *Opt. Express* **15**, 17172-17205 (2007).
9. Casimir, H. B. G. & Polder, D. The Influence of Retardation on the London-Vanderwaals Forces. *Physical Review* **73**, 360-372 (1948).
10. Lamoreaux, S. K. Demonstration of the casimir force in the 0.6 to 6 mu m range. *Phys. Rev. Lett.* **78**, 5-8 (1997).
11. Chan, H. B. Quantum mechanical actuation of microelectromechanical systems by the Casimir force. *Science* **293**, 1766-1766 (2001).
12. Munday, J. N., Capasso, F. & Parsegian, V. A. Measured long-range repulsive Casimir-Lifshitz forces. *Nature* **457**, 170-173 (2009).
13. Mizrahi, A. & Schachter, L. Two-slab all-optical spring. *Opt. Lett.* **32**, 692-694 (2007).
14. Feynman, R. P., Leighton, R. B. & Sands, M. L. *The Feynman lectures on physics*. Vol. I (Addison-Wesley, 1989).

15. Masmanidis, S. C. *et al.* Multifunctional nanomechanical systems via tunably coupled piezoelectric actuation. *Science* **317**, 780-783 (2007).
16. Kippenberg, T. J. & Vahala, K. J. Cavity optomechanics: Back-action at the mesoscale. *Science* **321**, 1172-1176 (2008).
17. LaHaye, M. D., Buu, O., Camarota, B. & Schwab, K. C. Approaching the quantum limit of a nanomechanical resonator. *Science* **304**, 74-77 (2004).
18. Wilson-Rae, I., Nooshi, N., Zwerger, W. & Kippenberg, T. J. Theory of Ground State Cooling of a Mechanical Oscillator Using Dynamical Backaction. *Phys. Rev. Lett.* **99**, 093901 (2007).
19. Thompson, J. D. *et al.* Strong dispersive coupling of a high-finesse cavity to a micromechanical membrane. *Nature* **452**, 72-75 (2008).